\documentclass[aip,sort,jcp]{revtex4}
\usepackage{amsmath}
\usepackage{subfigure}
\usepackage{graphicx}
\newcommand{\braket}[2]{\langle #1|#2\rangle}

\newcommand{\be}{\begin{equation}}
\newcommand{\ee}{\end{equation}}
\newcommand{\ba}{\begin{eqnarray}}
\newcommand{\ea}{\end{eqnarray}}
\def\ket#1{\left\vert #1 \right\rangle}
\def\bra#1{\left\langle #1 \right\vert}

\newcommand{\I}{\mathbf{1}}

\begin{document}

\author{Peter J. Love}
\affiliation{Department of Physics, Haverford College 370 Lancaster Ave. Haverford, PA 19041}
\title{Back to the Future: A roadmap for quantum simulation from vintage quantum chemistry}
\maketitle

\section{Introduction}
         
Quantum computation is the investigation of the properties of devices which use quantum mechanics to process information~\cite{Kitaev:book,Nielsen:book}.  Such quantum computers exist in the laboratory now, but currently are only able to process a few bits of information. Many experimental proposals for quantum computing exist - all of which share the goal of the controlled manipulation of the quantum state of the computer. These include superconducting systems~\cite{RevModPhys.73.357}, trapped atoms and ions~\cite{RevModPhys.71.S253,Wineland:98}, Nuclear Magnetic Resonance (NMR)~\cite{RevModPhys.76.1037}, optical implementations~\cite{RevModPhys.79.135,Knill:2001p5243}, Rydberg atoms~\cite{Saffman:2010p8002}, quantum dots~\cite{PhysRevA.57.120,qcqd}, electrons on Helium~\cite{Platzman:99} and other solid state approches~\cite{Kane:98,Kane:00}. For certain specialized applications it is believed that quantum computers are more powerful than their classical counterparts~\cite{Shor1997a,Grover:97,Abrams:97,Abrams:99,Kitaev:95}. 

One of the most promising areas of application for future quantum computers is quantum simulation: the use of a quantum computer to simulate or emulate another quantum system. Feynman's idea that quantum machines could be useful for simulating quantum systems was one of the founding ideas of quantum computing~\cite{Feynman:QC}. In the case of lattice Hamiltonians, specific techniques were first developed by Lloyd~\cite{Lloyd:96}. Zalka and Wiesner~\cite{Zalka:98,Wiesner} developed methods for wave mechanics and Lidar and Wang used this to define an algorithm for the calculation of the thermal rate constant~\cite{Wang}. Work to date in the area of quantum simulation has considered both static~\cite{Abrams:97} and dynamic properties of quantum systems~\cite{Zalka:98,Wiesner}. Recently these ideas have been applied to the quantum statics and dynamics of molecular systems~\cite{Aspuru-Guzik2005,guzikpnas}.

Accurate computation of the ground state energy of molecules is a basic goal in chemistry, but an area where accurate results remain expensive and elusive in many areas. This is also an area where quantum computers are known to offer significant advantages over classical machines for which the cost of numerically exact calculation grows exponentially with the size of the problem~\cite{Aspuru-Guzik2005}. In the context of physical chemistry quantum computers are suited to full-CI electronic structure calculations and to the simulation of chemical reactions {\em without} the Born-Oppenheimer approximation~\cite{Aspuru-Guzik2005,guzikpnas}. In applicable cases, these quantum algorithms provide an exponential performance advantage over classical methods. Work to date indicates that a quantum computer with a few hundred qubits would be a revolutionary tool for quantum chemistry. This number of qubits is a realistic target for sustained experimental development. However, given the small number of qubits currently available any application of quantum computing faces the challenge - how do we get there from here?

Quantum computation is slowly moving out of its infancy and it is now possible to consider the implementation of the smallest realizations of many of the basic algorithms known to present an advantage over classical computation. Significant experimental progress has been made in NMR~\cite{Negrevergne:2006p9474}, in trapped atoms and ions~\cite{Benhelm:2008p9154,Lanyon:2011p9111,Riebe:04,Barreiro:2011p9113,Monz:2011p9115}, superconductors~\cite{Mariantoni:2011p8990,Yamamoto:2010p9110,Johnson:2011p9482}, optical implementations of the gate model~\cite{OBrien:2004p9467,Politi:2009p9470,Marshall:2009p9471,Lanyon:2007p9393} and other approaches~\cite{Walther:2005p5266}. While the field may be maturing beyond its earliest days, there remains a large amount of work to be done before quantum simulation can compete with existing classical computers. How should we think about the early efforts to implement quantum algorithms for chemistry, and how much progress do they represent towards the goal of real quantum chemical discovery on quantum devices? This chapter proposes a framework for evaluating partial progress towards useful quantum simulators for quantum chemistry. The basic idea is to use the over sixty years of development of quantum chemical methods, implementations and results on classical machines to guide and evaluate the development of methods for quantum computers.

A great deal of effort was focussed on quantum chemistry over the last sixty years, and it is possible to go back to the literature and find the calculations performed on early computers - some from the vacuum tube era~\cite{SCH16}. These calculations are trivial by modern standards, but provide exemplars which could be implemented on early quantum computers (with, say, five to twenty quantum bits). By starting at this point, one can then trace the evolution of the classical calculations to the present day, and thereby lay out a roadmap leading from calculations which can be performed on quantum computers now to calculations at the research frontier. 

For calculations in which $1-20$ qubits are used to represent the wavefunction we propose a historical approach to the development of experimental realizations of quantum chemistry on quantum computers.  The first algorithm in this sequence has been implemented experimentally in linear optics quantum computing for a minimal basis set representation of the Hydrogen molecule~\cite{Lanyon2008Sub,Du:2010p6548}. 

We first address the merit of such an approach: Why not simply plug in appropriate minimal bases from modern electronic structure methods for simple systems? One motivation is to reexamine the problems of quantum chemistry from a new viewpoint.  In spite of the polynomial scaling of these algorithms with the problem size on quantum computers, gate and qubit estimates for simulation remain large.  Theoretical improvements to these algorithms could significantly reduce the experimental requirements for simulation of technologically interesting molecules. Methods adapted for classical computers may not be the best for quantum computers, and so looking back at methods which were perhaps abandoned is worthwhile. Another motivation is to avoid the temptation to focus on those problems that are particularly suited to quantum simulation, and to instead evaluate the methods against a pre-selected set of benchmarks. As such, these historical benchmarks provide a complement to valuable investigations of a set of standard chemical examples, as is performed in~\cite{pittner1}. Finally, it is interesting to measure progress in the quantum implementation against the historical development of the classical methods. This gives a straightforward way of estimating the distance yet to be covered before quantum techniques can compete with classical methods.

This example-by-example approach should be contrasted with more general theoretical inquiries. Much theoretical work in quantum information is focussed on quantum complexity theory~\cite{Cleve:99,bv}. In particular, the Local Hamiltonian problem has been extensively studied~\cite{Kitaev:book,Kempe:04,OsborneHC}. The Local Hamiltonian Problem formulates the task of finding the ground state of a Hamiltonian in the form of a decision problem. As such, the problem can be studied in terms of two solution concepts. First, one may consider versions of the problem where the solution may be found efficiently. Classically such examples would fall into the complexity classes P (Polynomial time) or, if a probabilistic algorithm is required, they would fall into BPP (Bounded-error Probabilistic Polynomial time). Secondly, one may consider those problems for which a given solution may be verified efficiently. Classically such problems would fall into the class NP (Non-Deterministic Polynomial time), or again, if the verification procedure is polynomial, into MA (Merlin-Arthur - see below for an explanation of this nomenclature). Of course, the question of whether the two solution concepts - finding versus verifying a solution - are in fact the same is the question of whether P=NP~\cite{PNP}.

In the classical world it is known that there are Hamiltonians for which verification of the ground state energy is as hard as any problem in NP~\cite{Barahona}. At least one classical algorithm for fermionic quantum systems, Quantum Monte Carlo, is also as difficult as any problem in NP in the worst case~\cite{PhysRevLett.94.170201}. On the quantum side, it is known that the eigenvalue decision problem in general for a local Hamiltonian on qubits is as hard as any problem in QMA~\cite{Kitaev:book,Kempe:04,OsborneHC}. However, these complexity-theoretic results are mostly for worst-case complexity, and hence one may state the advantage of quantum algorithms for ground state properties in complexity theoretic terms as dependent on the existence of a subset of physically interesting problems that are quantumly tractable but classically hard. 

In the present paper we do not propose an approach to obtaining complexity-theoretical results on the efficiency of quantum algorithms for ground state problems. Such a project might be described as Newtonian or Cartesian - the determination of grand theory covering all possible cases. Needless to say, such lofty goals are difficult to acheive in practice. Complexity theoretic proofs of the advantage of many widely used classical algorithms are few and far between. Indeed, in the case of Density Functional Theory and Quantum Monte Carlo the proofs are in the other direction, showing that the worst cases of these algorithms are likely to be hard, just as is the case for the local Hamiltonian problem~\cite{PhysRevLett.94.170201,Schuch:2009p5917,Kitaev:book,Kempe:04,OsborneHC}.

We may compare the Cartesian approach of quantum complexity theory to the approach of the present paper. We propose instead a Baconian approach, where specific examples are investigated in detail one at a time. These two approaches were recently contrasted by Dyson~\cite{BFDyson}, who characterized them in terms of Cartesian birds and Baconian frogs. The Cartesian birds fly over the landscape describing its broad features and general properties, whereas the Baconian frogs are happy to sit in the mud and investigate the details of what lies near them. Of course, we hope that these approaches are complementary. What we learn from the detailed investigation of a set of specific examples should be of use in correctly defining the right conditions under which quantum algorithms for ground state problems can be rigorously proved to be efficient. Setting theoretical considerations to one side we also hope that the elucidation of a well defined set of examples, and their use as a yardstick of progress, will help in the further development of experimental procedures and hardware for quantum simulation.

\section{Quantum Computing}

In this section we briefly review some salient techniques of quantum computation. Here we shall focus on brute force methods aimed at producing the minimal number of qubits - suitable for the first experimental implementations. Such implementations are useful for several reasons, not least because as they determine whether the experiments have sufficient precision to obtain the eigenvalues to a reasonable accuracy. For further details an the fully scalable approach we refer the reader to~\cite{Aspuru-Guzik2005} and to the Chapter in this volume by Aspuru-Guzik {\em et al.}, and~\cite{Whitfield:2010p6253,alanreviewII}. In those works the fully scalable approach to quantum simulation is reviewed in detail.

\subsection{Phase estimation}

\begin{figure}
\includegraphics[width = 0.7\textwidth]{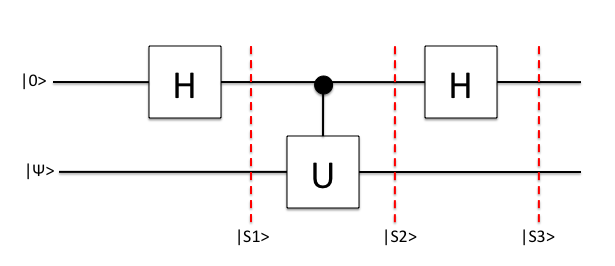}
\caption{A schematic circuit for one step of the phase estimation algorithm. The upper line is the readout register (taken to be one qubit here) and the lower line is the state register which carries an encoding of an eigenstate of the time evolution operator $U$.}
\label{Fi:fig1} 
\end{figure}

Phase estimation is a fundamental primitive of quantum computation~\cite{Kitaev:95,Kitaev:book,Nielsen:book}. Here we review those elements of the procedure that are relevant for our subsequent discussion. The circuit we shall analyze is shown in Figure~\ref{Fi:fig1}. It is composed of two registers - the readout register - which we take to be one qubit, and the state register, which we consider to be an arbitrary number of qubits.
The input state to the readout register is simply logical zero, $\ket{0}$, while the input state on the state register is an eigenstate of the Hamiltonian, and therefore of the time evolution operator:
\be
U\ket{\psi} = e^{-i\phi}\ket{\psi}
\ee
where 
\be
\phi = \frac{E\tau}{\hbar}
\ee
for evolution time $\tau$ and eigenenergy $E$. A Hadamard gate (which is the Fourier transform over the group $Z/2Z$) is applied to the readout register, resulting in state:
\be
\begin{split}
\ket{S1}&=\ket{+}\ket{\psi}\\
&= \frac{1}{\sqrt{2}}\biggl[\ket{0}\ket{\psi}+\ket{1}\ket{\psi}\biggr].\\
\end{split}
\ee
The next stage is a controlled application of the unitary operator $U$ corresponding to time evolution for a period $\tau$, resulting in state:
\be
\begin{split}
\ket{S2} &= \frac{1}{\sqrt{2}}\biggl[\ket{0}\ket{\psi}+e^{-i\phi}\ket{1}\ket{\psi}\biggr]\\
&=\frac{1}{\sqrt{2}}\biggl[\ket{0}+e^{-i\phi}\ket{1}\biggr]\ket{\psi}\\
\end{split}
\ee
The final step before measurement is to again apply the Hadamard gate to the readout qubit, resulting in the state:
\be
\ket{S3}= \frac{1}{2}\biggl[(1+e^{-i\phi})\ket{0}+(1-e^{-i\phi})\ket{1}\biggr]\ket{\psi}
\ee
It is worth noting that in the case where $\ket{\psi}$ is exactly equal to an eigenvector of $U$, the readout and state registers are not entangled at all in states $\ket{S1}$, $\ket{S2}$, $\ket{S3}$. A measurement performed on the readout qubit when the system is in state $\ket{S3}$ results in zero or one with the following probabilities:
\begin{equation}
\begin{split}
{\rm Prob}(0) &= |\frac{1+e^{-i\phi}}{2}|^2 = 1+\cos(\phi/2)\\
{\rm Prob}(1) &= |\frac{1-e^{-i\phi}}{2}|^2  = 1-\cos(\phi/2)\\
\end{split}
\end{equation}

We can now proceed from our description of phase estimation to energy measurement, following Abrams and Lloyd~\cite{Abrams1999a}. Suppose that we can estimate $\arccos(\langle \sigma_z\rangle/2) = \phi$ to fixed precision. Let $U$ be the time evolution operator of a system for time $t$:
\begin{equation}\nonumber
U = \exp(-it\hat H/\hbar)~~~~~~\phi = tE/\hbar 
\end{equation}
suppose the energy scale is such that $E/\hbar <1$ so that we can write $E/\hbar$ as a binary fraction:
\begin{equation}\nonumber
E/\hbar= 0.E_2E_4E_8E_{16}\dots = \frac{E_2}{2}+\frac{E_4}{4}+\frac{E_8}{8} +\frac{E_{16}}{16}+\dots
\end{equation}
then choose $t=2\pi 2^n$ so that:
\begin{equation}
\phi = N2\pi + \pi E_{2^n}/2 +\dots
\end{equation}
Using the recursive phase estimation procedure define in~\cite{Aspuru-Guzik2005,Whitfield:2010p6253,alanreviewII} we can repeat this calculation to estimate $E$ one bit at a time. This is the procedure that has been implemented for calculations of the Hydrogen molecule in~\cite{Lanyon2008Sub,Du:2010p6548}.

\subsection{Time evolution and the Cartan Decomposition}

Evidently, the procedure outlined above requires the implementation of controlled operators corresponding to time evolution for a sequence of times $2^i\tau$ for $i=1,2,\dots$. What is the difficulty of this problem if we know nothing about the form of $U$ except its dimension, i.e. all that is known is the number of qubits in the state register? The problem we wish to address is the decomposition of an arbitrary unitary operator into a sequence of one and two qubit gates. These one and two qubit gates are the basic primitive operations available to the experimentalist, and so we wish to proceed from  an arbitrary unitary to a quantum circuit over a particular gate set. A typical choice is the controlled not (CNOT) operation combined with arbitrary operations on single qubits - and this choice is known to be {\em universal}, i.e. sufficient to implement any unitary operator~\cite{PhysRevA.52.3457}. Much work in quantum computing has addressed this problem, including approaches that set bounds on the number of gates required~\cite{Glaser00a,shende:012310}, and approaches which can constructively obtain the full circuit, including all parameters for the one and two qubit gates~\cite{PhysRevA.52.3457,nakajima-2005,earp,drury2008}. For factoring an $q$ qubit unitary over CNOT gates and arbitrary one qubit rotations the best known performance is:
\be\label{bound}
\frac{23}{48}4^q -\frac{3}{2}2^q + \frac{4}{3}
\ee
CNOT gates and $4^q -1$ one qubit gates, where a one qubit gate is either an $X$, $Y$ or $Z$ rotation on a single qubit. 

The number of gates for an arbitrary unitary grows exponentially with the number of qubits. This is expected for a method that pays no attention to any special structure that may be present in $U$ beyond the number of qubits on which it acts and the tensor product structure of the Hilbert space of a set of qubits. Scalable implementations of time evolution operators must exploit the structure of the Hamiltonian in order to realize a polynomially scaling number of gates. Fortunately, for quantum chemical Hamiltonians in the second quantized formalism it is well understood how to do this, with a number of gates rising as $q^5$. For details of the scalable approach see the article by Aspuru-Guzik in this volume, or~\cite{Whitfield:2010p6253,alanreviewII}.

The bounds given in~\ref{bound} arises from a technique referred to as the quantum Shannon decomposition~\cite{shende:012310}. The underlying mathematical tool used in much of the literature is the Cartan Decomposition of a Lie Algebra~\cite{helgason,Carter}, and the corresponding decomposition for the Lie Group - in this case the unitary group.

What is the Cartan decomposition for a Lie group? Let us restrict to unitary operators of dimension $U(2^n)$ with determinant one, meaning the Lie group $SU(2^q)$. Let us call this group $G$, with the corresponding algebra being the antihermitian matrices (which physically correspond to Hermitian observables, times $i$). The first step in the decomposition is to identify a subalgebra, and hence a subgroup of G. We call this subgroup $K$. For our purposes we may consider two types of subalgebra (and hence subgroup). Firstly we can choose the subgroup to be the real unitary matrices, in other words the orthogonal group $SO(2^q)$, with the corresponding subalgebra being real antisymmetric matrices. Secondly we can consider block diagonal unitaries with blocks of size $p$ and $r$ such that $p+r=2^n$. It is natural for systems composed of qubits to choose $p=r=2^{q-1}$, so that the blocks are half the size of the original blocks. 

\begin{figure}[h]
\begin{center}
\subfigure[]{\includegraphics[width=1.7in]{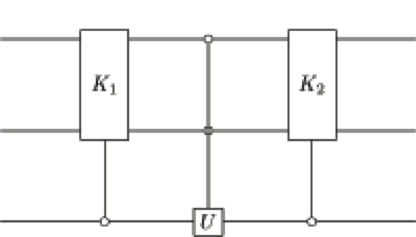}}
\subfigure[]{\includegraphics[width=3in]{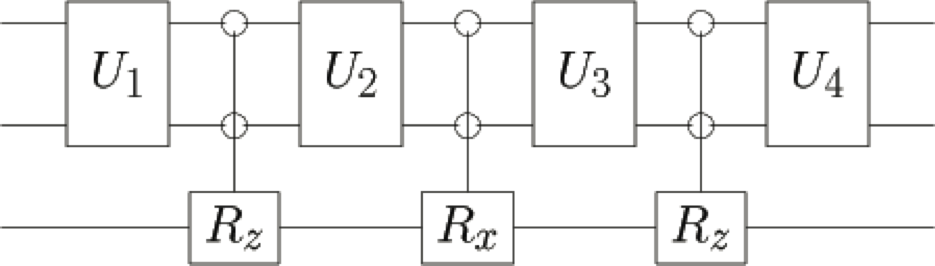}}
\end{center}
\caption{The two steps in the Quantum Shannon decomposition treated in~\cite{drury2008}, illustrated for three qubits. a) In the first stage of the decomposition the operator is broken into two operations controlled on the low qubit, separated by an operation on the low qubit controlled on the other qubits. b) The two outer controlled operations are themselves decomposed, resulting in four operations on two (in general, $n-1$) qubits. In the case of $n$ qubits, these $n-1$ qubit unitaries are factored again, until a circuit using only one and two qubit gates is obtained. This decomposition is the quantum Shannon decomposition~\cite{shende:012310}, and using Cartan involutions one can implement the decomposition given an arbitrary unitary matrix and produce the factors~\cite{drury2008}.}
\end{figure}

Having selected our subalgebra we have the algebra $g$ corresponding to the original group $G=SU(2^n)$, the subalgebra $k$ with corresponding subgroup $K$. The orthogonal complement of $k$ in $g$ is called $m$, and for a Cartan decomposition we find that the commutators obey the following relations. Firstly, $k$ is a subalgebra, so that:
\be
[k_1,k_2] \in k ~~{\rm for}~k_1~{\rm and}~k_2~{\rm in}~k.
\ee
Secondly, we have:
\be
[k_1,m_1] \in m ~~{\rm for}~k_1~{in}~k~{\rm and}~m_1~{\rm in}~m.
\ee
finally we find:
\be
[m_1,m_2] \in k ~~{\rm for}~m_1~{in}~{\rm and}~m_2~{\rm in}~m.
\ee
In particular, we note that any subalgebra of $m$ ($m$ itself is not a subalgebra) must be abelian, as all commutators must be zero, or they would lie in $k$, violating the subalgebra property. Hence there may be abelian subalgebras of $m$ and we may find such a subalgebra of largest dimension, call it $a$. It is a fact, that we shall not prove here, that any element of $m$ may be written:
\be
K_1^\dagger h K_1 = m_1,
\ee
where $K_1$ is an element of the subgroup $K$ and $h$ is an element of a maximal abelian subalgebra of $m$. Furthermore, every element of $G$ can be written:
\be
G_1 = K_2 M_1 = \exp(k_2)\exp{m_1},~~k_2\in k~~m_1\in m
\ee
and we may write:
\be
G_1 = \exp(k_2)\exp{m_1} = \exp(k_2)\exp{K_1^\dagger h K_1},~~k_2,k_1\in k~~m_1\in m,~~h\in a
\ee
using an elementary property of the exponential map we obtain:
\be
G_1 = \exp(k_2)\exp{K_1^\dagger h K_1} = \exp(k_2)K_1^\dagger \exp{h} K_1,~~k_2,k_1\in k~~m_1\in m,~~h\in a
\ee
which we may write:
\be
G_1 = K_3 A K_1 
\ee
where $K_3 = \exp(k_2)K_1^\dagger$  and $A=\exp(h)$. To put this in somewhat more physical language, given a Cartan decomposition of a unitary group one can find a maximal commuting subalgebra $h$ (i.e. a set of observables that may simultaneously take definite values) and a subalgebra $K$. Then the whole group can be written as a product of elements of $K$ and of $A$, the abelian group corresponding to the maximal abelian subalgebra.
For detailed proofs of the above assertions we refer the reader to the mathematics literature~\cite{helgason,Carter}.

So far we have only sketched the existence of this decomposition.  One further feature of these decompositions enables them to be obtained directly for a given element of $G$ and specific $K$ and $a$. This is the existence of a Cartan involution - a self inverse map that fixes the subalgebra $k$ and negates elements of the complement $m$:
\be
\begin{split}
\theta(k) &= k\\
\theta(m) &=  -m\\
\theta(\theta(x)) &= x\\
\end{split}
\ee
The involution is commutator preserving:
\be
\theta([x,y]) = [\theta(x),\theta(y)]
\ee
and these maps also extend to the group, with the following useful properties:
\be\label{inprop}
\Theta(K) = K,~~~\Theta(M) = M^{-1}, ~~~\Theta(KM) = KM^{-1}
\ee
Perhaps the most straightforward example is the involution for the orthogonal subalgebra, where:
\be
\theta(x) = -x^T,~~~~\Theta(X) = X^*
\ee
and the corresponding involution for the case where $K$ is the subalgebra of block diagonal matrices - for example where $p=q=2^{n-1}$:
\be
\theta(x) = \begin{pmatrix} I_{p\times p} &0\\0& - I_{r\times r}\end{pmatrix} x  \begin{pmatrix} I_{p\times p} &0\\0& - I_{r\times r}\end{pmatrix}
\ee
 where $I_{p\times p}$ is the $p\times p$ identity matrix. Similarly
\be
\Theta(X) =  \begin{pmatrix} I_{p\times p} &0\\0& - I_{r\times r}\end{pmatrix} X  \begin{pmatrix} I_{p\times p} &0\\0& - I_{r\times r} \end{pmatrix}
\ee
Given these involutions and their properties~\ref{inprop} it is straightforward to obtain $M^2$:
\be
M^2 = (MK^{-1})KM = \Theta(M^\dagger K^\dagger)KM = \Theta(G^\dagger)G
\ee
From this point all other factors may be obtained, and recursive application of a sequence of alternating decompositions can obtain the quantum shannon decomposition~\cite{drury2008,nakajima-2005}. 
 
The utility of this decomposition is many-fold. Firstly, for two qubit circuits the accidental isomorphism $SU(2)\times SU(2) = SO(4)$ means that the Cartan decomposition over the orthogonal subalgebra can be used to obtain circuits in which the $K$ factors are simply one qubit rotations~\cite{Makhlin:2002p8609,Zhang:2003p8004}. Secondly, it was recently shown that the quantum Shannon decomposition could be realized by using a pair of Cartan decompositions of the Lie algebra of the Unitary group~\cite{drury2008}. This insight provided a simple constructive method for obtaining circuits given a unitary operator. This method has also been implemented in the python programming language and extensively tested on random unitary matrices on up to 8 qubits. The method is suprisingly fast - a fit to timing data gives a rule of thumb timing of $T = 2^{2.5(q-4.5)}~~{\rm seconds}$ for a $q$ qubit unitary, corresponding to about a minute for a seven qubit unitary operator (a $128\times 128$ matrix).

\section{Quantum Chemistry - the CI method}

The starting point for problems in full-CI calculations is the electronic Hamiltonian expressed in a basis.  We briefly review the scalable second quantized approach, referred to as the Direct mapping in~\cite{Aspuru-Guzik2005}, which is covered extensively in the chapter in this volume by Aspuru-Guzik, before moving on to describe the Compact mapping of~\cite{Aspuru-Guzik2005}. The Compact mapping is based on a treatment of the CI matrix, and so we describe the construction of that matrix and the method for simulation arising from the use of the CI matrix. we close the section by comparing the direct and compact mappings for small numbers of qubits.  

\subsection{Second quantization - the Direct Mapping}

The Hamiltonian for the full quantum treatment of a set of nuclei and electrons is:
\begin{equation}
{\hat H}^{mol} = {\hat T_e}+ {\hat T_Z}+{\hat V}_{ZZ}(L_{pq})+{\hat V}_{ee}(r_{ij})+{\hat V}_{eZ}(R_{pi})
\end{equation}
where $\hat T_e$ and ${\hat T_Z}$ are the electron and nuclear kinetic energies, respectively,  ${\hat V}_{eZ}(R_{pi})$ and  ${\hat V}_{ee}(r_{ij})$ are the electron-nuclear and electron-electron coulomb interactions, and ${\hat V}_{ZZ}(L_{pq})$ is the nuclear-nuclear coulomb interaction. If we invoke the Born-Oppenheimer approximation and neglect $T_Z$ and treat $V_{ZZ}$ classically then we obtain:
\begin{equation}
{\hat H}^{elec} = -\frac{1}{2} \sum_{i=1}^N \vec{\nabla}_i^2  - \sum_{i,L} \frac{Z_L}{r_{iL}} + \sum_{i>j}^N \frac{1}{r_{ij}}
\end{equation}
which is the Hamiltonian for the electronic structure problem. 

To attack this problem via quantum computation, one requires a mapping of the time evolution operator into a set of unitary operations performed on one or two qubits (a quantum circuit). In the direct mapping this is accomplished by mapping the states of qubits into the occupation number basis. One therefore requires the same number of qubits as spin-orbitals, and a qubit in state one indicates that the corresponding spin-orbital is occupied. The Hamiltonian that acts on the fermionic states is:
\begin{equation}\label{2q}
H=\sum_{p,q}h_{pq}a^\dag_p a_q +\frac{1}{2}\sum_{p,q,r,s} h_{pqrs} a^\dag_p a^\dag_q a_r a_s,
\end{equation}
where $a^\dag_p$ is the creation operator for spin-orbital $p$ and $a_p$ is the annihilation operator for spin-orbital $p$.

The operator $a_j$ acts on basis vectors as follows:
\begin{equation}
\begin{array}{l}
\displaystyle
a_j\, \ket{l_{0},\ldots,l_{j-1},1,l_{j+1},\ldots,l_{q-1}}
\,=\, \bigl(-1\bigr)^{\sum_{s=0}^{j-1}\_s}\,
\ket{l_0,\ldots,l_{j-1},0,l_{j+1},\ldots,l_{q-1}} , \\
a_j\, \ket{lp_{0},\ldots,l_{j-1},0,l_{j+1},\ldots,l_{q-1}}
\,=\, 0;
\end{array}
\end{equation}
where $a^\dag_j$ is the Hermitian conjugate. In order to realize the Hamiltonian~\ref{2q} as an operator on the qubits we must implement the action of the creation and annihilation operators on the qubit states. We do so using the Jordan Wigner transformation following~\cite{Ortiz:2001p9489,Bravyi:2002p7994}. The creation operator is written in terms of Pauli matrices as 
$a^\dag_{qubit}$ = $\frac{1}{2}$($\sigma^x$ - $i\sigma^y$) = $\sigma^-$ , we can completely express the fermionic annihilation and
creation operators in terms of Pauli matrices:
\begin{equation}
\begin{array}{ll}
    a_j\equiv{\sigma^{z}}^{\otimes j-1}\otimes \sigma^{+} \otimes {\I}^{\otimes q-j}=\left[ \begin{array}{ll}
		1&0\\0&-1
	\end{array}\right]^{\otimes j-1}\otimes \left[ \begin{array}{ll}
		0&1\\0&0
	\end{array}\right]\otimes \left[ \begin{array}{rr}
		\phantom{+}1&0\\0&1
	\end{array}\right]^{\otimes q-j}\\
    a_j^\dagger\equiv{\sigma^z}^{\otimes j-1}\otimes \sigma^{-} \otimes {\I}^{\otimes q-j}=\left[ \begin{array}{ll}
		1&0\\0&-1
	\end{array}\right]^{\otimes j-1}\otimes \left[ \begin{array}{ll}
		0&0\\1&0
	\end{array}\right]\otimes \left[ \begin{array}{rr}
		\phantom{+}1&0\\0&1
	\end{array}\right]^{\otimes n-j}
\end{array}
\end{equation}
This correspondence is the Jordan-Wigner transformation. The use of this method imposes a number of extra gate applications required that scales as $O(q)$. As the general Hamiltonian~\ref{2q} has order $q^4$ terms the use of this method imposes a number of gates scaling at least as $q^5$, where $q$ is the number of qubits (equivalently of spin-orbitals). Because the terms in the Hamiltonian~\ref{2q} do not commute one may not simply separate the terms into gates. However, one may make use of the Trotter-Suzuki decompositions for small timesteps in order to obtain a circuit with polynomial number of gates. The number of gates in such decompositions scales as the number of terms in the Hamiltonian, with a prefactor depending on the details of the method, which in turn set the accuracy with which the circuit reproduces the time evolution operator. The generic scaling with number of qubits is therefore $q^5$ for these scalable methods.  

\subsection{FCI - the Compact mapping}

The alternative to second quantized is, of course, the first quantized approach in which all symmetry properties are retained in the wavefunction. We take as our starting point a set of $2K$ spin-orbitals (single electron basis functions), for a system of $N$ electrons. From these basis functions we may form 
\be
2K\choose N
\ee
antisymmetrized $N$ electron basis functions, or Slater determinants. It is convenient to classify these with reference to a Hartree-Fock basis state - the Slater determinant $\ket{\Psi_0}$ formed from the $N$ lowest energy spin-orbitals. One may label states by their differences from the Hartree-Fock state. The single excitations are labelled $\ket{\Psi_a^r}$, where spin-orbital $a$ in $\ket{\Psi_0}$ is replaced by spin-orbital $r$. Similarly, doubles and triples are labelled $\ket{\Psi_{ab}^{rs}}$, $\ket{\Psi_{abc}^{rst}}$ and so on. The number of $n$-tuple excitations is:
\be
{N\choose n}{2K-N\choose n}
\ee  
summing over $n$ gives the total number of Slater determinants:
\be
{2K\choose N}=\sum_{n=0}^N{N\choose n}{2K-N\choose n}
\ee
a result obtained from the Vandermonde convolution and the symmetry property of binomial coefficients.

Given this basis of determinants one wishes to obtain the Hamiltonian matrix elements within this basis - the CI matrix. To obtain the matrix elements of the Hamiltonian one uses Slater's rules~\cite{Slater:1931p8989,Slater:1929p6760,Szabo1996}. Given a matrix element $\bra{A}H\ket{B}$, where $A$ and $B$ are configurations (lists of spin-orbital labels), we may compute the CI matrix elements as follows:
\begin{enumerate}
\item{Diagonal elements. If $A=B$ so that the determinants are identical
\be
\bra{A}H\ket{A} = \sum_p^n \bra{p} \hat h\ket{p} +\sum_{p<q}^n \biggl( \braket{pq}{pq} - \braket{pq}{qp}   \biggr)
\ee}
\item{If the configurations $A$ and $B$ differ in one spin-orbital, $p$ in $A$ and $q$ in $B$
\be
\bra{A}H\ket{B} =  \bra{p} \hat h\ket{q} +\sum_{l}^n \biggl( \braket{pl}{ql} - \braket{pl}{lq}   \biggr)
\ee
}
\item{If the configurations $A$ and $B$ differ in two spin-orbitals, $pq$ in $A$ and $rs$ in $B$
\be
\bra{A}H\ket{B} =   \braket{pq}{rs} - \braket{pq}{sr}  
\ee
}
\item{If the configurations $A$ and $B$ differ in more than two spin-orbitals
\be
\bra{A}H\ket{B} =  0
\ee
}
\end{enumerate}
where we have defined the one-electron integrals:
\be
\bra{p} \hat h\ket{q} = -\frac{\hbar^2}{2m}\int \phi_p^*({\bf x}) \nabla^2\phi_q({\bf x}) d{\bf x}
\ee
and the two electron integrals:
\be
 \braket{pq}{rs}  = \int \phi_p^*({\bf x}_1) \phi_q^*({\bf x}_2) \frac{1}{|{\bf x}_1-{\bf x}_2|}\phi_r({\bf x}_1) \phi_s({\bf x}_2) d{\bf x}_1 d{\bf x}_2.
\ee

At this stage quantum simulation of FCI via the compact mapping comes into focus. The compact mapping is a mapping of the logical states of the qubits onto labels of Slater determinants. The matrix eigenvalue problem we wish to solve is defined by the CI matrix, and we implement time evolution according to this Hamiltonian by, for example, using a Cartan decomposition of the relevant time evolution operator. Ironically, while FCI is typically the classical method that uses the largest number of states for a given problem, in the quantum case the compact mapping uses a smaller Hilbert space dimension than the direct mapping of the second quantized approach. The reason for this is that the second quantized approach maps the entire Fock space of the system into the Hilbert space of the qubits, whereas the compact mapping only maps the $N$ electron sector into the quantum computer.

The procedure outlined above for the compact mapping is not scalable. Although the input data to the CI matrix is the same set of one and two electron integrals used to define the second quantized Hamiltonian, the compact mapping does not respect the tensor product structure of the qubit Hilbert space, and hence there is no known procedure analogous to the Trotter-Suzuki decomposition that enables the efficient construction of circuits with a number of gates rising polynomially in the number of spin-orbitals. In addition, while for the second quantized formalism the one and two electron integrals appear directly in the circuit as rotation angles for quantum gates acting on a small number of qubits, at present there is no known way of constructing the time evolution operator corresponding to the CI matrix on the quantum computer without computing it classically first.

The compact mapping remains of interest for a number of reasons, however. If we consider problems in which we have $2K$ spin-orbitals, and therefore the second quantized approach requires $2K$ qubits (for up to $2K$ electrons), the compact mapping requires:
\be
c(k,N) = \biggl\lceil \log_2{2k \choose N}\biggr\rceil 
\ee
which is less than $2K$ for all $N<2K$, being the nearest whole number of qubits into which the $N$ particle sector of the full fock space of up to $2K$ particles will fit. Hence the compact mapping will always enable the mapping of larger problems than the direct mapping into the qubit Hilbert space.

This advantage is only of interest if one can in fact take advantage of the smaller qubit requirements by implementing the time evolution operator. Evidently, the non-scalable nature of the compact mapping makes it asymptotically uninteresting. However, the issue for small examples is not quite so clear cut. Suppose the number of gates required to implement the evolution operator which yields the $U$ is $A(2K)^5$, where $A$ is a constant. In order to obtain $p$ bits of precision in the final answer one must run at least one phase estimation procedure with $2^pA(2K)^5$ gates. The corresponding numbers for performing the Cartan involution are much worse for the first stage of iterative phase estimation, where a Cartan involution requires 
\be
2^{2c(k,N)} =  {2k \choose N}^2
\ee
gates, as well as a comparable amount of classical computation to obtain the circuit. However, each stage of the iterative phase estimation requires only the same number of gates, as one simply computes new parameters in the Cartan decomposition for each evolution operator $U^{2^p}$. Hence the total gate count comparison is roughly:
\be\label{esteq1}
G_{\rm Compact} \simeq p {2k \choose N}^2 
\ee
compared with
\be\label{esteq2}
G_{\rm Direct} = 2^p A (2K)^5
\ee 

\begin{figure}[h]
\begin{center}
\includegraphics[width=0.45\textwidth]{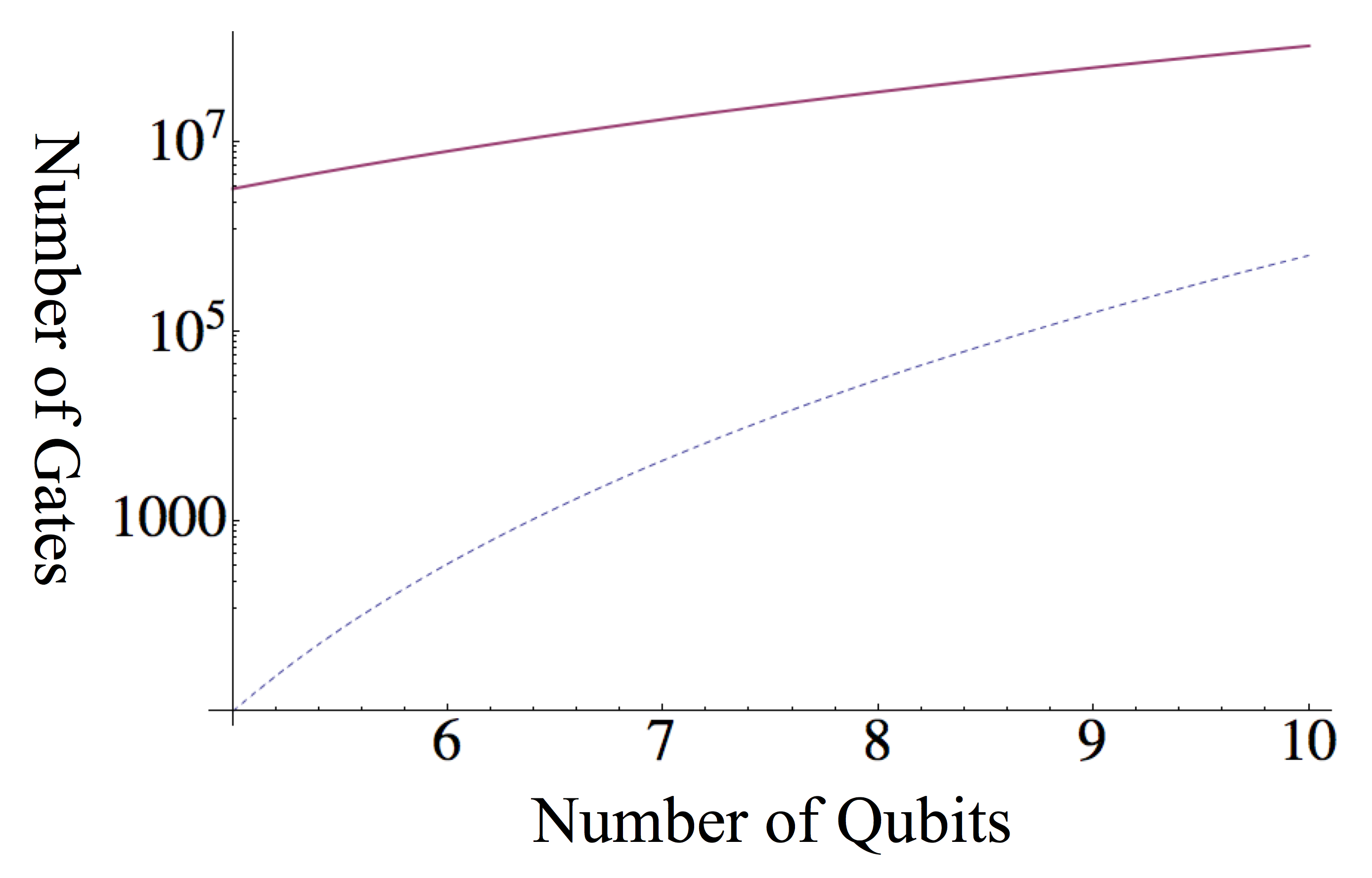}
\includegraphics[width=0.45\textwidth]{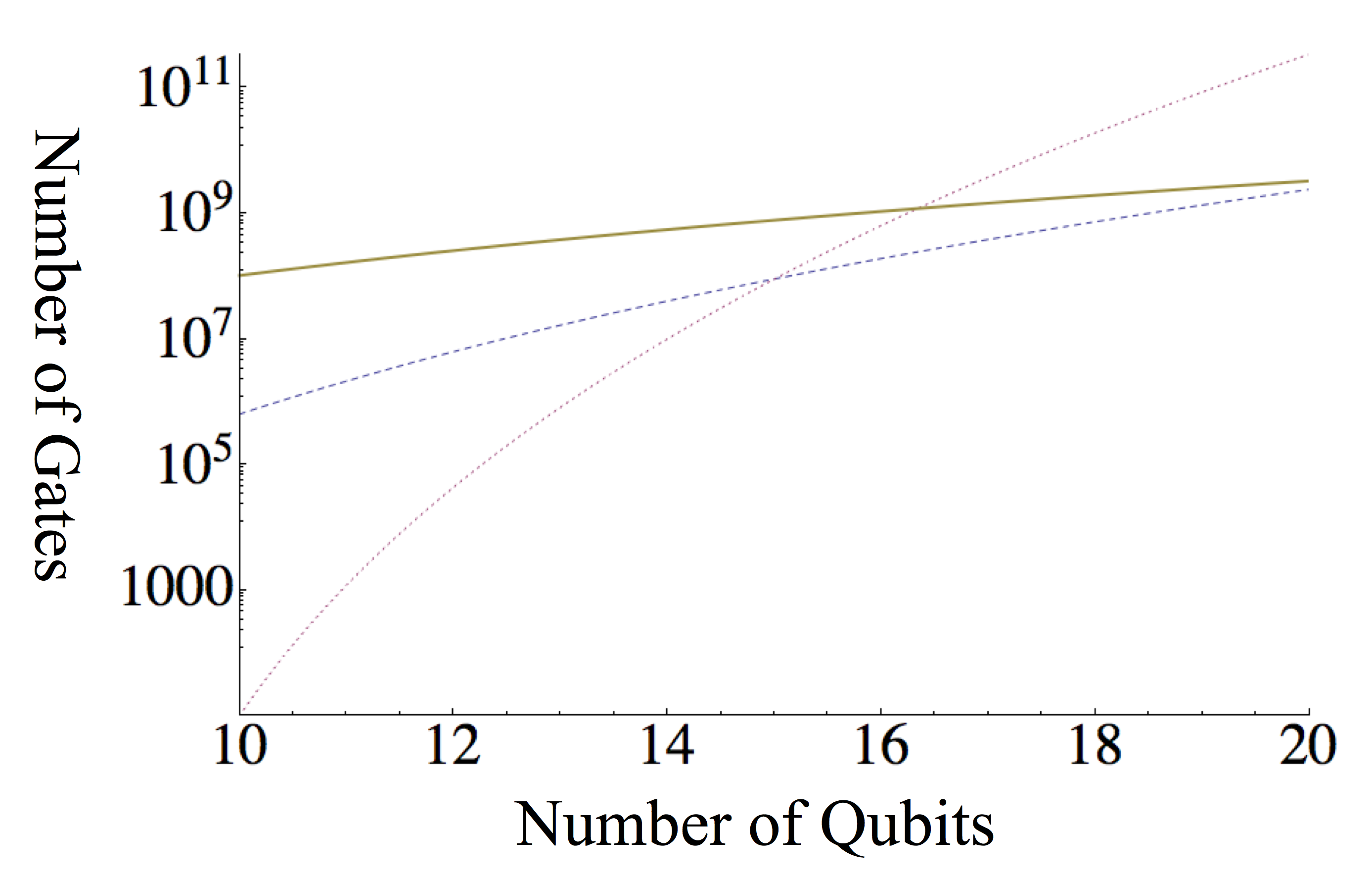}
\caption{Left: A comparison of gate counts, as estimated by equations~(\ref{esteq1}) and~(\ref{esteq2}) (with $A=1$) for $5$ electrons and numbers of qubits between $5$ and $10$. Solid line: $G_{\rm Direct}$. Dashed line, $G_{\rm Compact}$. This indicates that for fewer than $5$ qubits the Compact mapping is preferred over the scalable Direct mapping. Right: A comparison of gate counts, as estimated by equations~\ref{esteq1} and~\ref{esteq2} (with $A=1$) for numbers of qubits between $5$ and $10$. Solid line: $G_{\rm Direct}$, $5$ electrons. Dashed line, $G_{\rm Compact}$, $5$ electrons. Dotted line, $G_{\rm Compact}$, $10$ electrons.  This indicates that the crossover between the  Compact mapping and the scalable Direct mapping lies between $10$ and $20$ qubits.}
\label{compfig1}
\end{center}
\end{figure}

The crossover point at which the advantage in number of gates of the direct mapping outweighs the advantage in qubits of the compact mapping represents a significant milestone for quantum simulation - the point at which the advantage of scalable methods is realized in practice. We plot $G_{\rm compact}$ and $G_{\rm Direct}$ in Figure~\ref{compfig1} for $A=1$, for $5$ and $10$ electrons and qubit numbers between $5$ and $20$. These estimates indicate that the crossover point between Compact and Direct mappings is between $10$ and $20$ qubits. 

\section{A Selection of Historical Calculations in Quantum Chemistry}

Here we present a sequence of calculations spanning from the 1930's to the 1960's. Our intent here is not to give a historically faithful account of the development of quantum chemistry. We merely choose a few mileposts and evaluate the cost in qubits of the Compact mapping defined above of those calculations on quantum hardware. The choice of these calculations was strongly influenced by the presentation of~\cite{Schaefer:Book}. For translations of earlier foundational papers in quantum chemistry we refer the reader to Hinne Hettema's volume~\cite{Hettema:Book}, which also includes a fascinating preface discussing the historical context of early work on quantum chemistry in both Germany and the United States. 

\subsection{The thirties and forties}

Perhaps the first exact ab initio calculation was performed by Coulson on a Brunsviga mechanical calculator for the molecular ion $H_3^+$ ~\footnote{For a biography of Coulson see the address given by Roy McWeeny, Professor of Theoretical Chemistry at University of Sheffield, which opened the Fifth Canadian Symposium on Theoretical Chemistry, being held at University of Ottawa in June, 1974, and was dedicated to the memory of Charles Coulson, F.R.S.}. Early calculations of the ground state energy of the Hydrogen molecule were conducted in the 1930's using the variational principle~\cite{SCH02}. The first studies by Heitler and London were superceded by the methods of Hyelleras in which a variational Ansatz is constructed from symmetry considerations alone. The calculation of the variational energy in these calculations naturally involved integrals over the Ansatz functions. 

\subsection{The fifties}

In 1950 Boys introduced the use of Gaussian functions~\cite{SCH06}. In the same year Boys gave a presentation of the CI method and presented a CI calculation of the ground state energy of Beryllium~\cite{SCH07}. Interestingly, Boys first CI calculation used Slater-type LCAO orbitals rather than Gaussians because of the relatively simpler integrals which occur in this atomic problem. He used three $s$ and one $p$ single-electron functions and performed the calculation using six and ten determinants formed from these functions. Here we see perhaps the first application of CI, corresponding to a three or four qubit problem.

In 1952 Taylor and Parr applied the CI method to the Helium atom, a system which they note is ``no longer an interesting problem {\em per se}'' because of the accuracy of Hyelleraas' variational solution obtained in 1930. However they note that in spite of the melancholy prospects for progress of the CI method Helium is the first non-trivial example. Their calculation involved four configurations $1s1s'$, $2p^2$, $3d^2$, $4f^2$ and hence corresponds to a two qubit calculation. Their paper reproduces $88\%$ of the correlation energy of Helium~\cite{SCH10}. Three years later, Shull and L\"owdin reconsidered the Helium atom using natural orbitals~\cite{SCH13}. They obtained the CI solution using six configurations from three orbitals (1s, 2s, 3s). The results were then analyzed in terms of natural orbitals, and they showed that similar results could be obtained with only three configurations. By doing so they reduced a three-qubit (six configuration) calculation to a two-qubit (three configuration) calculation.
         
In 1956 CI calculations were reported on BH with 7 orbitals, 23 co-detors, and 3 internuclear distances, $H_2O$ with 8 orbitals, 96 co-detors, at seven nuclear configurations. These results were obtained by Boys using the EDSAC computer at Cambridge~\cite{SCH16}. The BH calculations are five-qubit problems, and the $H_2O$ calculations are seven qubit problems. The following year Miller, Friedman, Hurst and Matsen continue the extension of CI to molecular problems with calculations of $LiH$ and $BeH^+$~\cite{SCH17}. Their calculations used 20 terms corresponding to all possible singlet states of $1s$, $2s$, $2p_\sigma$ Slater orbitals on the metal atoms and $1s$ on the Hydrogen. They performed six, ten and twenty determinant calculations on IBM CPC and IBM 650 machines, these are three, four and five qubit problems respectively. In the same year, Pekeris gave a new method for the determination of the ground state energy of two-electron atoms~\cite{SCH18}.

In a largely philosophical paper published in 1959 Mulliken and Roothaan reflected on the status of exact quantum chemistry noting that ``it so happens that the more difficult problems of chemistry and molecular physics are of a high order of complexity in terms of the mathematical analysis and the computational efforts required for quantitative results. However, and this is the point of our paper, those who are working in this field are now catching up, and we believe that the results of theoretical computations are going to compete more and more strongly with experiment from now on''~\cite{SCH20}. 

\begin{table}
  \centering 
\begin{tabular}{|c|c|c|c|}
	\hline
Year   & Calculation & Citation  & Number of qubits\\
	\hline
1933   & $H_2$ & \cite{SCH02} & $1$\\
	\hline
1950   & $Be$ & \cite{SCH07} & $3,4$\\
	\hline
1952   & $He$ & \cite{SCH10}& $2$\\
	\hline
1955   & $He$ & \cite{SCH13}& $2,3$\\
	\hline	
1956   & $BH$ & \cite{SCH16}& $5$\\
	\hline	
1956   & $H_2O$ & \cite{SCH16}& $7$\\
	\hline
1957   & $LiH$ & \cite{SCH17}& $3,4,5$\\
	\hline
1957   & $BeH^+$ & \cite{SCH17}& $3,4,5$\\
	\hline
1960   & $Be$ & \cite{SCH23}& $6$\\
	\hline
1960   & $CH_2$ & \cite{SCH28}& $19$\\
	\hline	
1963   & $H_2$ & \cite{SCH38}& $3,4,5,6$\\
	\hline
1966   & $HeH$ & \cite{SCH54}& $3$\\
	\hline
1966   & $Li_2$ & \cite{SCH54}& $3$\\
	\hline
1967 &$H_2O$ & \cite{SCH66ix} &$10$\\
	\hline
1967 &$H_2O$ & \cite{SCH66xiv} &$24$\\
	\hline
1967 &$H_2O$ & \cite{SCH66x,SCH66xii} &$38, 39$\\
	\hline
1968 &$H_2O$ & \cite{SCH66} &$39,46$\\
	\hline
1968   & $Be$ & \cite{SCH67}& $11$\\
	\hline
1969   & $Li$, $Be^+$, $B^{++}$ & \cite{SCH72}& $9,10$\\
	\hline
1969   & $BH$, $FH$ & \cite{SCH75}& $12,14$\\
	\hline
1970   & $H_2O$ & \cite{SCH76}& $23$\\
	\hline
\end{tabular} 
\vspace{0.5cm}
  \caption{A sequence of quantum chemical problems and their minimal qubit requirements. These calculations bridge the period of development of CI from the first variational calculations of the hydrogen molecule energy to the advent of the GAUSSIAN-70 program. The qubit costs are associated with the Compact mapping of full-CI calculations on the system studied in the cited publication, although the methods used originally may not be full-CI.}\label{CIVintage}
\end{table}
          
\subsection{The sixties}
In 1960  Watson published a CI study of $Be$ involving $37$ configurations constructed from a basis of STO's- a six qubit problem~\cite{SCH23}. Boys and Cook also lay out the progress made in developing the CI method over the previous decade in~\cite{SCH27}. The following year Sinanoglu published perhaps the first in depth study of electron correlation, introducing the idea that the correlation energy would be well approximated by the sum of pair correlation energies~\cite{SCH29}.

In 1963 Barnett at MIT reports on the development of a FORTRAN code for the IBM 709/90 which implements SCF and CI calculations on basis sets of size $<50$ although no actual example calculations apart from integral evaluations are reported in this work~\cite{SCH37}. In the same year Hagstrom and Shull report on the efficacy of natural orbitals for $H_2$ and $H_2^*$. They report that calculations using $21$ and $33$ configurations may be replaced by natural orbital bases of size $8$ and $15$ respectively~\cite{SCH38}. For our purposes this is of particular interest as it converts five and six qubit calculations into three and four qubit calculations.  Natural orbital calculations for $HeH$ and $Li_2$ followed in 1966 providing seven configuration (or three qubit) calculations~\cite{SCH54}. These calculations are of interest as the natural orbital wavefunctions provide a class of ansatz states whose complexity of preparation can be investigated.

In the same year (1968) Bunge published a new calculation of the ground state energy of the $Be$ atom, using a $180$ term CI with $1492$ Slater  determinants - an eleven qubit calculation~\cite{SCH67}. A study in 1969 of the $\pi\pi^*$ states of the ethylene molecule compared HF and CI calculations~\cite{SCH69}. An improved CI algorithm was applied to three-electron problems ($Li$, $Be^+$ and $B^{++}$) using 7 to 9 Slater orbitals, (9 and 10 qubit problems respectively)~\cite{SCH72}. Calculations on $H_3$ were also performed with $9$ Slater orbitals, another $10$ qubit problem. The publication by Bender and Davidson in 1969 of a study of the first row diatomic hydrides using very large CI provides a rich set of examples~\cite{SCH75}. Two example calculations are of $BH$, a twelve qubit calculation, and of FH, a fourteen qubit calculation.

In 1970, a new study of water with a $14$ function basis set, corresponding to a 23 qubit calculation, comparing STO and contracted Gaussian basis sets~\cite{SCH76}. In the same year one of the first calculations using the GAUSSIAN-70 code appeared - treating 30 species with STO-3G basis sets in which Slater type functions are approximated by three Gaussians~\cite{SCH80}. With this paper and the advent of codes such as GAUSSIAN which achieved wide usage in the theoretical quantum chemistry community full CI calculations became common. The Pople classification of basis sets makes the determination of a number of qubits for any large CI calculation straightforward, and the number of qubits is typically $20-50$ for calculations which historically lie between 1970 and the present. These basis sets and relevant citations are readily available from, e. g.~\cite{BSE}. 

 \begin{table}
  \centering 
\begin{tabular}{|c|c|c|c|c|c|}
	\hline
Calculation   & Method &  Size of basis & Energy (a.u.) & Number of qubits\\
\hline
Reeves and Boys~\cite{SCH66xiii}& MC STO  & $\dots$ &-75.776 & $\dots$\\
\hline
McWeeny and Ohno~\cite{SCH66ix}& MC STO & $7$ &-75.761 & $10$\\
\hline
Moccia& OC STO  & $28$ &-75.992 & $36$\\
\hline
Harrison~\cite{SCH66xiv}& MC GLF  & $14$ &-76.002 & $24$\\
\hline
Moskowitz and Harrison~\cite{SCH66xii}& MC GTO  & $36$ &-76.034 & $39$\\
\hline
Ritchie and King~\cite{SCH66x}& MC CGF  & $38$ &-76.034 & $38$\\
\hline
\cite{SCH66}& MC GLF  & $32$ &-76.044 & $39$\\
\hline
\cite{SCH66}& MC GTO & $56$ &-76.002 & $46$\\
\hline
\end{tabular} 
\vspace{0.5cm}
  \caption{SCF calculations of the water molecule ground state energy in 1967 and 1968. These calculations, aiming to approach the Hartree-Fock limit for water define large molecular orbital bases. The qubit numbers refer to the minimal requirements to perform a full-CI in the molecular orbital basis defined by the published SCF calculation. Unsuprisingly, SCF calculations which approach the Hartree-Fock limit define full-CI problems requiring several tens of qubits.}\label{waterscf}
\end{table}

In 1960 several papers applying variational and LCAO-SCF methods to molecules appeared. Mclean reports LCAO-SCF on $C_2H_2$ and $CO_2$ with molecular orbitals constructed from (for $C_2H_2$) 1s STO's on each $H$ and $1s$, $2s$, $2p\sigma$ and $2p\pi$ STO's on each $C$. For $CO_2$ MO's are formed from $1s$, $2s$, $2p\sigma$ and $2p\pi$ STO's on each $O$ and $1s$, $2s$, $2p\sigma$ and $2p\pi$ on the $C$~\cite{SCH21}. In the same year there was also a calculation of the $^1\Sigma^+_g$ excited state of $H_2$ by a variational method~\cite{SCH22}.

 In 1960 Nesbet reports an LCAO Hartree-Fock calculation on HF using $9 \sigma$ and $5\pi$ orbitals using the IBM 704 computer~\cite{SCH26}.  Foster and Boys report a study of the three lowest lying states of $6$ different bond angles for the $CH_2$ radical using a basis of six exponential functions on the carbon and one on each hydrogen~\cite{SCH28}. A CI calculation in this basis would represent a $19$ qubit problem.

Much work in the previous century was devoted to approaching the Hartree-Fock limit using ever better SCF methods. For the water molecule, an example of this work is given by~\cite{SCH66}. In this work, progress to date (prior to 1968) for the SCF treatment of water is summarized in Table IX, which we reproduce in Table~\ref{waterscf}. These SCF calculations provide an interesting reference: repeating these calculations using CI yields more challenging, but still tractable problems. The data in Table~\ref{waterscf} give us qubit requirements in the range $10-46$
  
\section{Boys' 1950 calculation for Be}

In this paper Boys constructs  a $12\times 12$ CF Hamiltonian over a set of configuration state functions (CSF's) which are referred to as ``co-detors'' - spin-adapted linear combinations of determinants. There are four sets of functions $sA$, $sB$, $sC$, and $pA$, where the $s$ and $p$ notation indicates that the functions in these sets are eigenvalues of the square of total angular momentum with quantum number $l=0$ (s) and $l=1$ (p). The letters $A$, $B$, $C$ denote different radial dependence, and for each radial dependence and total orbital angular momentum there is one function for each possible combination of the eigenvalues of $L_z$ and $S_z$. Because these are atomic orbitals (single electron eigenfunctions) the total spin eigenvalue is $1/2$. Writing out the elements of each set we obtain:
\be
\begin{split}
sA&=\{sA(0,-1/2), sA(0,1/2)\}\\
sB&=\{sB(0,-1/2), sB(0,1/2)\}\\
sC&=\{sC(0,-1/2), sC(0,1/2)\}\\
pA&=\{pA(-1,-1/2), pA(-1,1/2),pA(0,-1/2), pA(0,1/2),pA(1,-1/2), pA(1,1/2)\}\\
\end{split}
\ee
Hence there are $12$ atomic orbitals here. However, Boys uses linear combinations of functions in this basis which are simultaneous zero eigenvalues of $\hat L^2$ and $\hat S^2$. He does so by taking the four electrons and placing them in two pairs, each pair being in a spin and orbital singlet.

The orbital coupling can use the special case of the Clebsch-Gordan coefficients relevant to the singlet state of total orbital angular momentum, $\braket{l_1,l_2,m_1,m_2}{J,M}$, where $l_1,l_2,m_1,M_2$ are the angular momentum quantum numbers labeling the orbitals and $J$ and $M$ are the total angular moment quantum numbers. By the recoupling rules for angular momentum such a state can only be produced from a pair of angular momentum states with equal values of $l$:
\be
\braket{l,l,m_1,m_2}{0,0} = \delta_{m_1,-m_2}\frac{(-1)^{l-m_1}}{\sqrt{2l+1}}.
\ee
This implies that the combination of orbital products is a uniform superposition of pairs with opposite $z$ components with alternating signs. 
\be
\begin{split}
\ket{0,0} &= \sum_{m_1 =-l}^l\sum_{m_2=-l}^l \delta_{m_1,-m_2}\frac{(-1)^{l-m_1}}{\sqrt{2l+1}}\ket{l,l,m_1,m_2}\\
&= \sum_{m_1 =-l}^l\frac{(-1)^{l-m_1}}{\sqrt{2l+1}}\ket{l,l,m_1,-m_1}\\\end{split}
\ee

These are denoted:
\be
\begin{split}
(sAsA) {^1S} (sBsB) {^1S},\\ 
(sAsA) {^1S} (sBsC) {^1S},\\  
(sAsA) {^1S} (pApA) {^1S},\\
(sAsB) {^1S} (pApA) {^1S}\\
\end{split}
\ee
These are then antisymmetrized by applying the operator:
\be
{\mathcal S} = \sum_{\pi\in S_4}{\rm sign}{\pi} \frac{\pi}{\sqrt{4!}}
\ee
Resulting in the normalized wavefunctions given in Eqn 4 in~\cite{SCH07}.

The elementary integrals (overlaps and expectation values for the one-electron orbitals) whose radial parts are given by:
\be
\braket{r}{A,l,a} = r^{A+l}e^{-ar}
\ee
These radial functions are not orthogonal. Their overlaps are given by:
\be
\braket{A,l,a}{B,l,b} = \int_0^\infty r^{A+B+2l}e^{-(a+b)r}   r^2dr = \frac{(A+B+2l+2)!}{(a+b)^{A+B+2l+3}}
\ee
The values of these parameters for the functions $sA$, $sB$, $sC$ and $pA$ are given in table~\ref{rad}.
\begin{table}[h]
  \centering 
\begin{tabular}{|c|c|c|c|c|c|c|c|c|}
	\hline
Orbital   & $\ket{A,l,a}$ & $\braket{r}{A,l,a}$  & $A$&$l$&$a$ &$\braket{\theta,\phi}{l,m}$&$m$\\
	   \hline
$sA$ &$\ket{0,0,4}$ & $e^{-4r}$ & $0$ &$0$ &$4$&$\sqrt{\frac{1}{4\pi}}$&$0$\\
$sB$ &$\ket{1,0,1}$ & $re^{-r}$ & $1$ &$0$ &$1$&$\sqrt{\frac{1}{4\pi}}$&$0$\\
$sC$ &$\ket{0,0,3}$ & $e^{-3r}$ & $0$ &$0$ &$3$&$\sqrt{\frac{1}{4\pi}}$&$0$\\
$pA$ &$\ket{0,1,1}$ & $re^{-r}$ & $0$ &$1$ &$1$&$\sqrt{\frac{3}{8\pi}}\sin\theta e^{-i\phi}$&$1$\\
$pA$ &$\ket{0,1,1}$ & $re^{-r}$ & $0$ &$1$ &$1$&$\sqrt{\frac{3}{8\pi}}\cos\theta$&$0$\\
$pA$ &$\ket{0,1,1}$ & $re^{-r}$ & $0$ &$1$ &$1$&$\sqrt{\frac{3}{8\pi}}\sin\theta e^{i\phi}$&$-1$\\
\hline
\end{tabular} 
\vspace{0.5cm}
  \caption{Hydrogenic orbitals for the calculation of~\cite{SCH07}}\label{rad}
\end{table}

\begin{table}[h]
  \centering 
\begin{tabular}{|c|c|c|c|c|c|}
	\hline
Orbital   & $\ket{A,l,a}$ & $\ket{0,0,4}$  & $\ket{1,0,1}$&$\ket{0,0,3}$\\
	   \hline
$sA$ &$\bra{0,0,4}$ & $(1/16)^2$ & $3!/5^4$ &$2!/7^3$ \\
$sB$ &$\bra{1,0,1}$ & $-$ & $4!/2^5$ &$3!/4^4$ \\
$sC$ &$\bra{0,0,3}$ & $-$ & $-$ &$2!/6^3$ \\
\hline
\end{tabular} 
\vspace{0.5cm}
  \caption{Overlap integrals for the calculation of~\cite{SCH07}}\label{over}
\end{table}

The three $p$ orbitals are orthogonal to the $s$ orbitals and to each other by virtue of the orthogonality of the angular parts of the wavefunction. The three $s$ orbitals are not orthogonal, but an orthogonal basis may be constructed from them by the Gram-Schmidt procedure. The corresponding orthonormal orbitals are:
\be
\begin{split}
\ket{\phi_1} &= \frac{1}{\sqrt{\braket{0,0,4}{0,0,4}}}\ket{0,0,4} = 16\ket{0,0,4}\\
\ket{\phi_2} &= \ket{1,0,1} - \frac{\braket{0,0,4}{1,0,1}}{\braket{0,0,4}{0,0,4}}\ket{0,0,4}\\
\ket{\phi_3} &= \frac{1}{\sqrt{\braket{\psi_3}{\psi_3}}}2^9\left[\frac{3}{5^4}\frac{5^47^3-2^{17}}{2^5 5^47^3-3 2^{17}7^3/5^4}-\frac{1}{\sqrt{\braket{\psi_3}{\psi_3}}}\frac{1}{7^3}\right]\ket{0,0,4}\\
&- \frac{1}{\sqrt{\braket{\psi_3}{\psi_3}}}\frac{5^47^3-2^{17}}{2^5 5^47^3-3\times2^{17}7^3/5^4}\ket{1,0,1}+\ket{0,0,3}\\
\ket{\phi_4}&=\sqrt{\frac{4}{3}}\ket{0,1,1}\\
\end{split}
\ee
Where the coefficients in eqn (18) in~\cite{SCH07} can now be seen to arise as follows:
\be
\begin{split}
\frac{1}{\sqrt{\braket{\psi_2}{\psi_2}}} &= 1.173302451\\
\frac{1}{\sqrt{\braket{\psi_2}{\psi_2}}} \frac{\braket{0,0,4}{1,0,1}}{\braket{0,0,4}{0,0,4}}=\frac{3!}{\sqrt{\braket{\psi_2}{\psi_2}}}\left(\frac{4}{5}\right)^4 &= 2.883508103\\
\frac{1}{\sqrt{\braket{\psi_3}{\psi_3}}}2^9\left[\frac{3}{5^4}\frac{5^47^3-2^{17}}{2^5 5^47^3-3 2^{17}7^3/5^4}-\frac{1}{\sqrt{\braket{\psi_3}{\psi_3}}}\frac{1}{7^3}\right] &= -69.59716965\\
\frac{1}{\sqrt{\braket{\psi_3}{\psi_3}}}\frac{5^47^3-2^{17}}{2^5 5^47^3-3\times2^{17}7^3/5^4}&=0.5968868832\\
\frac{1}{\sqrt{\braket{\psi_3}{\psi_3}}}&=47.60738095\\
\sqrt{\frac{4}{3}}&=1.154700538\\
\end{split}
\ee

After all that, the six by six CI matrix actually computed by Boys was:

\be
\begin{pmatrix}
0.& 0.3967&0.2488& -0.9416& -0.1361& 0.\\
0.3967 &17.0628& -0.1257& -0.5601& 0.& 0.\\
0.2488& -0.1257& 5.0387&  0.1499& -0.0099&0.0196\\
-0.9416& -0.5601& 0.1499& 8.7275&  0.& -0.1361\\
-0.1361& 0.& -0.0099& 0.& 0.3677& -0.9502\\
0.& 0.&  0.0196& -0.1361& -0.9502& 9.1242\\
\end{pmatrix}
\ee
\

\begin{table}
  \centering 
\begin{tabular}{|c|c|c|}
\hline
Boys Designation & Qubit state &\\
\hline
${\mathcal A}(sAsA){^1S}(sBsB){^1S}$ & $\ket{0}$&\\
${\mathcal A}(sBsB){^1S}(sCsC){^1S}$ & $\ket{1}$&\\
${\mathcal A}(sAsA){^1S}(sBsC){^1S}$ & $\ket{2}$&\\
${\mathcal A}(sAsC){^1S}(sBsB){^1S}$ & $\ket{3}$&\\
${\mathcal A}(sAsA){^1S}(pApA){^1S}$ & $\ket{4}$&\\
${\mathcal A}(sAsC){^1S}(pApA){^1S}$ & $\ket{5}$&\\
${\mathcal A}(sAsA){^1S}(sCsC){^1S}$ & $\ket{6}$&\\
${\mathcal A}(sAsB){^1S}(sCsC){^1S}$ & $\ket{7}$&\\
${\mathcal A}(sBsB){^1S}(pApA){^1S}$ & $\ket{8}$&\\
${\mathcal A}(sAsB){^1S}(pApA){^1S}$ & $\ket{9}$&\\
\hline
\end{tabular} 
\vspace{0.5cm}
  \caption{Compact mapping of Boys basis for Beryllium}\label{boyscompact}
\end{table}

\begin{table}
\begin{tabular}{|c|c|c|c|c|c|c|c|c|c|c|}
\hline
&$\ket{0}$&$\ket{1}$&$\ket{2}$&$\ket{3}$&$\ket{4}$&$\ket{5}$&$\ket{6}$&$\ket{7}$&$\ket{8}$&$\ket{9}$\\
\hline
$\ket{0}$ &$-14.4577$&$0.3967$&$0.2488$&$-0.9416$&$-0.1361$&$0.$&$0.0541$&$0.0986$&$-0.1361$&$0.0276$\\
$\ket{1}$ &$0.3967$&$2.6051$&$ -0.1257$&$-0.5601$&$0.$&$ 0.$&$0.0220$&$-0.1958$&$-0.0317$&$0$\\
$\ket{2}$ &$0.2488$&$-0.1257$&$-9.419$&$0.1499$&$-0.0099$&$0.0196$&$0.3534$&$1.0192$&$0$&$0.0070$\\
$\ket{3}$ &$-0.9416$&$-0.5601$&$0.1499$&$-5.7302$&$0.$&$-0.1361$&$0.0269$&$-0.8085$&$0$&$-0.1780$\\
$\ket{4}$ &$-0.1361$&$0.$&$-0.0099$&$0.$&$-14.09$&$-0.9502$&$-0.0132$&$0$&$0.0220$&$-0.2816$\\
$\ket{5}$ &$0.$&$0.$&$0.0196$&$-0.1361$&$-0.9502$&$-5.3335$&$0.0215$&$0.0070$&$-0.0269$&$0.0407$\\
$\ket{6}$ &$0.0541$&$0.0220$&$0.3534$&$0.0269$&$-0.0132$&$0.0215$&$-3.7904$&$1.3259$&$0$&$0$\\
$\ket{7}$ &$0.0986$&$-0.1958$&$1.0192$&$-0.8085$&$0$&$0.0070$&$1.3259$&$1.5002$&$0$&$-0.0946$\\
$\ket{8}$ &$-0.1361$&$-0.0317$&$0$&$0$&$0.0220$&$-0.0269$&$0$&$0$&$-4.1937$&$0.0032$\\
$\ket{9}$ &$0.0276$&$0$&$0.0070$&$-0.1780$&$-0.2816$&$0.0407$&$0$&$0.0946$&$0.0032$&$-10.1121$\\
\hline
\end{tabular} 
\vspace{0.5cm}
  \caption{CI matrix for Boys basis for Beryllium}\label{boyscompact}
\end{table}\label{BoysCI}

Given the CI matrix and established a compact mapping for the co-detors we can consider a non-scalable implementation of Boys original calculation. Our starting point is the $6\times6$ version of the CI matrix. As noted above, the task is to decompose powers of the evolution operator corresponding to this Hermitian matrix into a quantum circuit. This may be accomplished by a Cartan decomposition that treats the evolution operator as an action on a qubit and a qutrit. While it is conventional to decompose unitaries over qubits, controlled quantum systems employing larger numbers of states than two have also been utilized in a number of experimental realizations~\cite{Lanyon:2008p9391}. A more conventional approach would utilize three qubits in the state register with two three qubit states uninvolved in the calculation. The optimal three-qubit circuit requires $44$ CNOT gates and $63$ one qubit gates~\cite{drury2008}. 

\section{Conclusion}

Elucidation of small cases of quantum chemical calculations from the past literature provides an interesting experimental road map leading from calculations which quantum computers can perform now, to calculations beyond the capabilities of comparable classical methods. Scalable methods for the calculation of ground state properties using quantum computers have been studied extensively in the last few years. The known scalable techniques for implementing time evolution operators of fermionic systems, comprising the Jordan Wigner transformation and Trotter-Suzuki decompositions, provide polynomial performance. However, they are most likely not yet optimized and further study of these scalable approaches is likely to be fruitful. At the other extreme, the approach detailed here considers the smallest calculations where non-scalable approaches that minimize the required number of qubits are used. These allow for early experimental implementation, the purpose of which is to demonstrate sufficient quantum control and coherence to obtain results of chemical accuracy for small systems. As we have argued here, there is a crossover point between scalable and non-scalable implementations which will likely result in two experimental frontiers. At one frontier as much chemistry as possible will be crammed into the available qubit Hilbert space as possible, whether the approach used is scalable or not. At the other frontier we will be proceeding with the development of the scalable methods that we know will carry us beyond model systems and ultimately beyond the capabilities of any conceivable classical computing device.

\section*{Acknowledgements}

The author would like to acknowledge productive discussions with Alan Aspuru-Guzik, James Whitfield, Ken Brown and Jake Seeley. This work is supported by National Science Foundation awards CHE-1037992 and PHY-0955518.


\end{document}